\documentclass[%
 reprint,
superscriptaddress,
nofootinbib,
 amsmath,amssymb,
 aps,
]{revtex4-2}

\usepackage{graphicx}
\usepackage{dcolumn}
\usepackage{bm}
\usepackage{hyperref}

\usepackage{physics}
\usepackage{siunitx}

\begin{document}

\preprint{APS/123-QED}

\title{Machine Learning-Informed Scattering Correlation Analysis of Sheared Colloids}
\author{Lijie Ding$^\dagger$}
\affiliation{Neutron Scattering Division, Oak Ridge National Laboratory, Oak Ridge, TN 37831, USA}
\author{Yihao Chen$^\dagger$}
\email{ychen258@sas.upenn.edu}
\affiliation{Department of Physics and Astronomy, University of Pennsylvania, Philadelphia, PA 19104, USA}

\author{Changwoo Do}
\affiliation{Neutron Scattering Division, Oak Ridge National Laboratory, Oak Ridge, TN 37831, USA}

\date{\today}

\begin{abstract}
We carry out theoretical analysis, Monte Carlo simulations and Machine Learning analysis to quantify microscopic rearrangements of dilute dispersions of spherical colloidal particles from coherent scattering intensity. Both monodisperse and polydisperse dispersions of colloids are created and undergo a rearrangement consisting of an affine simple shear and non-affine rearrangement using Monte Carlo method. We calculate the coherent scattering intensity of the dispersions and the correlation function of intensity before and after the rearrangement, and generate a large data set of angular correlation functions for varying system parameters, including number density, polydispersity, shear strain, and non-affine rearrangement. Singular value decomposition of the data set shows the feasibility of machine learning inversion from the correlation function for the polydispersity, shear strain, and non-affine rearrangement using only three parameters. A Gaussian process regressor is then trained based on the data set and can retrieve the affine shear strain, non-affine rearrangement, and polydispersity with a relative error of 3\%, 1\% and 6\%, respectively. Together, our model provides a framework for quantitative studies of both steady and non-steady microscopic dynamics of colloidal dispersions using coherent scattering methods.
\end{abstract}

\maketitle
\def\thefootnote{$\dagger$}\footnotetext{Equal contribution. Y.C. conceived this work and carried out theoretical analysis. L.D. carried out Monte Carlo simulations and Machine Learning analysis.}\def\thefootnote{\arabic{footnote}}

\section{Introduction}
Quantification of the microscopic dynamics of materials made of nano- to micrometer-scale constituents is vital in understanding the origins of macroscopic mechanical properties and designing novel functional materials for pharmaceutical, environmental and other industrial applications\cite{wu2020mechanical}. While traditional optical microscopy can provide real space information, it is limited by its resolution and the opacity of the materials\cite{badon2017multiple}. Scattering techniques\cite{murphy2020capillary,guinier1955small} like x-ray photon correlation spectroscopy\cite{chu2001small, shpyrko2014x, leheny2015, madsen2020structural} (XPCS), dynamic light scattering\cite{goldburg1999dynamic, aime2019probing} (DLS) and small-angle neutron scattering\cite{shibayama2011small,chen1986small} (SANS) provide great opportunities to probe the microscopic information of such materials, and they were deployed to study the micro-structural dynamics of colloidal~\cite{chen2020, donley2023}, polymeric~\cite{ruocco2013}, atomic materials~\cite{luttich2018anti}. However, the challenge of scattering techniques is the quantification of microscopic rearrangement in real space back from the scattering patterns (Fourier space), which for most of the time are only available in a limited range of 2-dimension (in case of an area detector) and sometimes even 1-dimension (in case of photon counting device). Previous efforts have investigated steady shear~\cite{burghardt2012x}, diffusion~\cite{leitner2009atomic}, localization of particles~\cite{chen2020microscopic}, etc. Most of the schemes characterize temporal correlation function of scattering intensity and require an average of the intensity or/and the correlation function of intensity in a time interval assuming a steady dynamics. On the other hand, non-steady microscopic dynamics, like non-affine rearrangement, are widely observed in real space and plays an important role in the non-linear properties of soft materials~\cite{keim2015role, wen2012non}, like yielding~\cite{jana2019correlations} and memory efforts~\cite{galloway2022relationships}. However, they are much less studied in coherent scattering experiments due to their non-linear and transient nature, and the interpretation often requires intensive modeling and computation~\cite{ma2014unveiling, he2024transport}. 

To address these issues, we develop a generalized theoretical and Machine Learning\cite{murphy2012machine,carleo2019machine} (ML) framework to quantify affine and non-affine rearrangements of dilute, both monodisperse and polydisperse, colloidal dispersions based on the correlation function of coherent scattering intensity. Our ML approach utilizes Monte Carlo\cite{krauth2006statistical} (MC) simulation to generate particle configurations and rearrangements in 2D. Coherent scattering intensity of the particles and the correlation function of the intensity before and after the rearrangement were calculated, and we extracted three essential features of the correlation function using singular value decomposition that reliably recover the magnitude of both affine and non-affine rearrangements and polydispersity. Similar approach have been applied to other soft matter systems including: colloids\cite{chang2022machine,huang2023model,tung2022small}, lamellar\cite{tung2024unveiling}, and polymers\cite{tung2023inferring,ding2024machine,ding2024machine_ladderpolymer}.  We then use Gaussian process regression\cite{williams2006gaussian} (GPR) to obtain a mapping from scattering data to the system parameters including polydispersity, shear strain, and non-affine rearrangement. We also tested our trained GPR using simulation data aside from the training data, good agreement between the ML extracted system parameters and the MC references were achieved, showing good accuracy of our approach. Our model can be easily adopted to coherent scattering experiments to extract microscopic rearrangements between two scattering patterns, which is especially useful for studies of non-steady and transient dynamics.

The rest of this paper is organized as the following: in Sec.~\ref{sec:method} we introduce our colloidal systems, the theoretical analysis of coherent scattering intensity, MC simulation, and a brief summary of the GPR. We then present our results in Sec.~\ref{sec:results}, where we illustrate the scattering intensity and correlation function of the colloidal dispersions under rearrangements, validate the feasibility for ML inversion of system parameters using singular value decomposition of the correlation, and show the application of ML analysis for scattering data using GPR. Finally, we summarize our paper and discuss potential future directions following this work in Sec.~\ref{sec:summary}.

\section{Method}
\label{sec:method}

\subsection{Coherent scattering and rearrangement transformation}

The normalized scattering intensity of a polydisperse dispersion of $N$ spherical particles with a configuration of 2-dimensional coordinate $\mathcal{S}=\left\{ (x,y)\right\}$ is given by\cite{chen1986small}:

\begin{equation}
    I(\vb{q};\mathcal{S}) = \frac{\left(\sum_{i=1}^{N} V_i F_i(\vb{q}) e^{-i \vb{q}\cdot \vb{r}_i}\right)\left(\sum_{i=1}^{N} V_i F_i^\dagger(\vb{q}) e^{i \vb{q}\cdot \vb{r}_i}\right)}{\sum_{i=1}^{N} V_i^2}
    \label{equ: Iq}
\end{equation}
where $V_i=\frac{4\pi}{3}R_i^3$ is the volume of particle $i$ with radius $R_i$, $\vb{q}$ is the scattering wave vector and $\vb{r}_i=(x_i,y_i)$ is the position of particle $i$. $F_i(\vb{q})$ is the form factor amplitude of the $i$th particle such that\cite{guinier1955small}:

\begin{equation}
    F_i(\vb{q}) = 3\frac{\sin(qR_i) - qR_i \cos(qR_i)}{(qR_i)^3}
\end{equation}
where $q=|\vb{q}|$ is the magnitude of the scattering wave vector. 

The rearrangement transformation of the particle positions consists of an affine simple shear deformation along $x$-axis (shear gradient runs along $y$-axis) with a shear strain $\gamma$, as shown in Fig.~\ref{fig:config_Iq}(a), and a non-affine rearrangement where the particle are randomly displaced by $\delta x_i$ and $\delta y_i$, which follow a Gaussian distribution with zero mean and a standard deviation $D_2$. Such a transformation $\Gamma$ is expressed as:
\begin{equation}
    \Gamma \mqty(x_i \\ y_i) = \mqty(x_i + \gamma y_i + \delta x_i \\ y_i + \delta y_i).
    \label{equ: transformation}
\end{equation}

In homodyne scattering, the average translation of all the particles has no effect on the scattering intensity, so we choose the form of Eqn.~\eqref{equ: transformation} to have a fixed origin for the affine shear.

The correlation function of two scattering intensities before and after the transformation $\Gamma$ is
\begin{equation}
    g(\vb{q}) = \frac{\left< I(\vb{q};\mathcal{S})I(\vb{q};\Gamma(\mathcal{S})) \right>_{\mathcal{S}}}{ \left< I(\vb{q};\mathcal{S})\right>_{\mathcal{S}}^2},
\end{equation}
where $\left<\dots\right>_{\mathcal{S}}$ is the average over different realizations of the configuration.

In case of dilute dispersions of monodisperse spherical particles, including materials doped with a dilute amount of monodisperse tracer particles, the positions of different particles and, thus, their rearrangements are uncorrelated, and the correlation function $g(\vb{q})$ associated with the transformation $\Gamma$ is predicted as (similar to that in ref.~\cite{aime2019probing}): 
\begin{equation}
    g(\vb{q}) = 1 + \text{sinc}^2(\frac{q\gamma L\cos\theta}{2})\exp(-q^2D^2_2),
    \label{equ:g_prediction}
\end{equation}
where $\text{sinc}(x) = \frac{\sin(x)}{x}$, and $\theta$ is the angle between $\vb{q}$ and $x$-axis.

\subsection{Monte Carlo simulation}
We sample the positions of particles $\mathcal{S}$ in a $[-L,L]^2$ square with a number density $n$. The radius of the particles follows a log-normal distribution such that $\ln R_i \sim \mathcal{N}(\ln R_0, R_s)$, where the polydispersity index of the dispersion, PDI = $\left<V_i^2\right>_i/\left<V_i\right>_i^2 = \left<R_i^6\right>_i/\left<R_i^3\right>_i^2 = \exp(9R_s^2)$, is controlled by $R_s$~\cite{kotz2019continuous}.

We calculate the scattering intensity $I(\vb{q})=I(q_x,q_y)=I(q\cos(\theta),q\sin(\theta))$ in the polar coordinates for all the particles inside the box of $[-0.5L,0.5L]^2$ before and after the $\Gamma$ transformation, then calculate the correlation function $g(\vb{q})$.  The values of $I(\vb{q})$ and $g(\vb{q})$ are averaged over $10^4$ samples of $\mathcal{S}$ for each set of system parameters $(n L^2, R_s,D_2,\gamma L)$ . We also calculate the radial and angular average of the correlation function such that $g(q) = \left<g(\vb{q}) \right>_{\theta}$ and $g(\theta) = \left<g(\vb{q}) \right>_{q}$, where $\left<\dots\right>_{\theta}$ and $\left<\dots\right>_{q}$ are averages over all measured $\theta$ and $q$, respectively. Without lose of generality, we use natural unit $R_0 = 1$ for the size of particles and the beam size $L=800R_0$. We measured the $I(\vb{q})$ and $g(\vb{q})$ respect to 100 different values of $q\in[0.5,5]$ uniformly distributed on a log scale, and 101 different values of $\theta\in[0,\pi]$ uniformly distributed on a linear scale, and we note that $I(q,\theta+\pi) = I(q,\theta)$ due to $\pm\vb{q}$ symmetry of Equ.~\eqref{equ: Iq}. The choice of $R_0$, $L$, and the range of $q$ is to mirror experimental conditions: a colloidal dispersion of particles with a radius of 10 nm, a synchrotron x-ray beam size of 8 $\mu$m, and an area detector at small angle scattering setup that covers a wave vector ranging from 0.05 to 0.5 nm$^{-1}$.

\subsection{Gaussian process regression}
To obtain the inverted mapping from scattering correlation function $\vb{x} = g(\vb{q})$ to system parameters, or inversion targets $\vb{y} = (n L^2, R_s,D_2,\gamma L)$, we train a GPR using the data generated by MC simulation. In the context of GP, the prior on the regression function is a Gaussian process $g(\vb{x})\sim GP(m(\vb{x}),k(\vb{x},\vb{x}'))$, where $m(\vb{x})$ is the prior mean function and $k(\vb{x},\vb{x}'))$ is the covariance function or kernel. The goal of the GPR is to find the optimized the posterior $p(\vb{Y}_*|\vb{X}_*,\vb{X},\vb{Y})$ of the function output $\vb{y}$. The joint distribution of Gaussian process is\cite{williams2006gaussian}:
\begin{equation}
    \mqty(\vb{Y}~\\ \vb{Y}_*) \sim \mathcal{N}\left( \mqty[m( \vb{X}~) \\ m(\vb{X}_*)], \mqty[k(\vb{X}~,\vb{X}~) & k(\vb{X}~,\vb{X}_*) \\ k(\vb{X}_*,\vb{X}~)~ & k(\vb{X}_*,\vb{X}_*)]  \right)
    \label{equ: Gaussian_process}
\end{equation}
where a constant function is used for the prior mean $m(\vb{x})$, and the kernel is consisted of a Radial basis function and a white noise. $k(\vb{x},\vb{x}') = e^{\frac{-|\vb{x} - \vb{x}'|^2}{2l}} + \sigma \delta(\vb{x},\vb{x}')$, where $\delta$ is the Kronecker delta function. $l$ and $\sigma$ denote the hyperparameters corresponding to the correlation length and variance of observational noise, which can be obtained by training on simulation data. In practice, we use the scikit-learn Gaussian Process library\cite{pedregosa2011scikit} for convenience of implementation and efficiency.

To investigate the distribution of $\vb{X}$, we define the pair distance distribution function (PDDF) for the $\vb{X}$:
\begin{equation}
    p(z) = \frac{1}{M^2}\sum_{i,j=1}^{M} \delta( |\vb{x}_i - \vb{x}_j| - z)
    \label{equ: pddf}
\end{equation}
where $M=|\vb{X}|$ is the number of data points. In addition, to help determining the proper range of hyperparameter $l$ when initiating then optimization process, we calculate the autocorrelation function (ACF) for feature $\mu$ \cite{chang2022machine}:
\begin{equation}
    C_{\mu}(z) = \frac{\left< \mu(\vb{x}) \mu(\vb{x}+z) \right>_{\vb{x}} - \left< \mu(\vb{x})\right>^2_{\vb{x}} }{\left< \mu^2(\vb{x})\right>_{\vb{x}} - \left< \mu(\vb{x})\right>^2_{\vb{x}}}
    \label{equ: ACF}
\end{equation}
where the $\left< \dots \right>_{\vb{x}}$ is average over all data point in $\vb{X}$.

\section{Results}
\label{sec:results}
\subsection{Scattering function of the dispersion}
\begin{figure}[!h]
    \centering
    \includegraphics{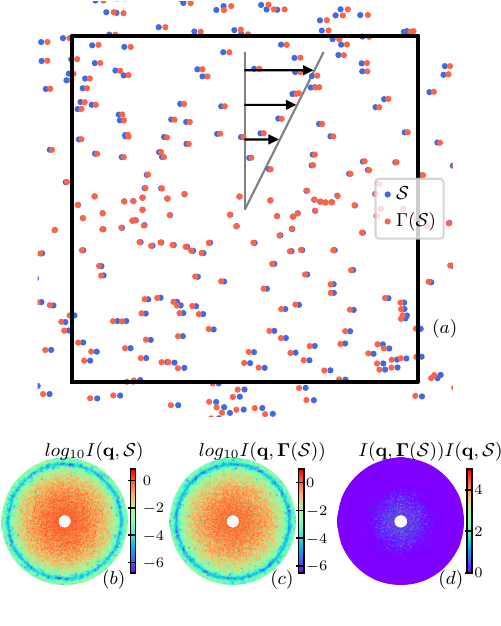}
    \caption{Illustration of a single configuration of particles undergoing transformation $\Gamma$, in which $(nL^2,R_s,D_2,\gamma L)=(150,0,0.5,30)$. Back frame indicate the region that beam line shines on. (a) Spatial distribution of particles before ($\mathcal{S}$) and after ($\Gamma(\mathcal{S})$) the transformation. For better visualization, the size of the particles are not in scale. Scattering intensity of the configuration (b) before and (c) after the transformation $\Gamma$. (d) Correlation function between the scattering intensity of transformed and non-transformed configurations.} 
    \label{fig:config_Iq}
\end{figure}

Figure~\ref{fig:config_Iq}(a) shows an example of the configuration $\mathcal{S}$ of monodisperse particles before and after the rearrangement transformation $\Gamma$, where $\gamma L$ = 50 and $D_2$ = 5. Figures~\ref{fig:config_Iq}(b) and (c) shows the corresponding coherent scattering patterns of configurations $\mathcal{S}$ and $\Gamma(\mathcal{S})$, respectively. Speckles are clearly seen in the scattering patterns. Figure~\ref{fig:config_Iq}(d) shows the product of the two instances of scattering intensity shown in Figs.~~\ref{fig:config_Iq}(b) and (c).

\begin{figure}[!h]
    \centering
    \includegraphics{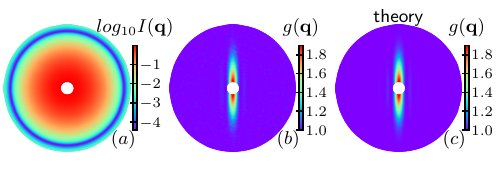}
    \caption{Scattering function $I(\vb{q})$ and scattering correlation function $g(\vb{q})$ of the particles with $(nL^2,R_s,D_2,\gamma L)=(150,0,0.5,10)$. The range of scattering wave vector is $q = |\vb{q}|\in [0.5, 5]$, plotted in linear scale. (a) Averaged scattering intensity. (b) Averaged correlation function. (c) Theoretical predicted correlation function for the monodisperse system as in Equ.~\eqref{equ:g_prediction}.}
    \label{fig:Iq_gq}
\end{figure}

Figure.~\ref{fig:Iq_gq}(a) shows an example of scattering pattern $I(\vb{q})$ of monodisperse particles after averaging over $20,000$ configurations of $\mathcal{S}$. The scattering intensity $I(\vb{q})$ is isotropic and reflects the form factor $F^2(\vb{q})$ of the spherical particles. Figure~\ref{fig:Iq_gq}(b) shows the correlation $g(\vb{q})$ averaged over $20,000$ samples. $g(\vb{q})$ is highly anisotropic with a high correlation in $y$-axis ($\theta = \pi/2$). The non-affine rearrangement is isotropic and causes the decay of structural correlation in all directions. For the affine shear rearrangement (Equ.~\eqref{equ: transformation}), the $y$-coordinates of particles remain unchanged, so the affine shear only leads to the decay of correlation in other directions except of $y$-axis (gradient of the affine shear). The pattern of $g(\vb{q})$ is captured very well by the theoretical prediction of Equ.~\eqref{equ:g_prediction} as shown in Fig.~~\ref{fig:Iq_gq}(c).

\begin{figure}[!h]
    \centering
    \includegraphics{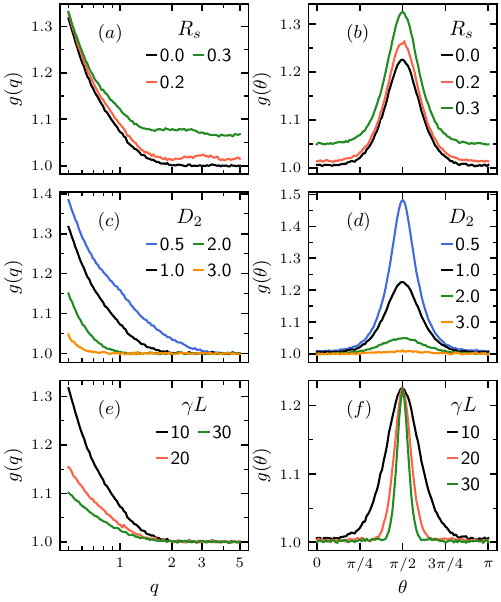}
    \caption{Radial (left column) and angular (right column) correlation function, $g(q)$ and $g(\theta)$, for various polydispersity $R_s$,  non-affine rearrangement $D_2$, and  affine shear $\gamma L$ with a reference of $(n,R_s,D_2 ,\gamma L)=(150,0,1,10)$ represented by black line. (a) and (b) are for various $R_s$, (c) and (d) are for various $D_2$, and (e) and (f) are for various $\gamma L$.}
    \label{fig:gq}
\end{figure}

To further quantify the effect of polydispersity, shear strain, and non-affine transformation on the correlation function,  we use $(R_s, D_2, \gamma L) = (0,1,10)$ as a baseline to demonstrate the effect of three system parameters $(R_s, D_2, \gamma L)$ on the $g(q)$ and $g(\theta)$. Fig.~\ref{fig:gq} shows the radial and angular averaged correlation function $g(q)$ and $g(\theta)$ with different values of $(R_s, D_2, \gamma L)$, which alter the correlation function in different ways. The radial correlation function $g(q)$, peak at the lowest scattering wave vector $q$, and decays at higher $q$. The variation of particle size $R_s$ controls the height of the plateau of the $g(q)$ at high q limit, which increases with larger $R_s$. Affine shear strain $\gamma L$ and non-affine rearrangement $D_2$ affect the peak of $g(q)$ at the lowest $q$ in a similar way where both increasing shear strain $\gamma L$ and $D_2$ lower the peak. The effect of the three system parameters on angular correlation function $g(\theta)$ is more distinguishable as shown in Figs.~\ref{fig:gq}(b), (d) and (f). The angular correlation function $g(\theta)$ has a peak at $\theta= \pi/2$ (affine shear gradient direction), where $\gamma L$, $D_2$, and $R_s$ affect the width, height, and baseline of the peak. Therefore, we focus on the angular correlation function for the rest of the work.

\subsection{Feasibility for Machine Learning inversion}
We generate a data set of $6,000$ angular scattering correlation functions $\vb{F} = \left\{g(\theta)\right\}$ whose corresponding system parameters $\vb{Y}=\left\{(nL^2,R_s,D_2,\gamma L)\right\}$ are randomly distributed such that: $nL^2\in U(100,200)$, $R_s \in U(0,0.3)$, $D_2 \in U(0.5,3)$ and $\gamma L \in U(5,30)$, where $U(a,b)$ is the uniform distribution in the interval $[a,b]$. Note that the $g(\theta)$ is measured for $101$ $\theta$ uniformly in $[0,\pi]$, so the $\vb{F}$ is a $6,000\times 101$ matrix. Following a similar framework in references\cite{chang2022machine, ding2024machine}, we carry out a principle component analysis of the data set matrix $\vb{F}$ using singular value decomposition (SVD) $\vb{F}=\vb{U}\vb{\Sigma}\vb{V}^T$, where $\vb{U}$, $\vb{\Sigma}$, and $\vb{V}$ are matrices of $6,000\times6,000$, $6,000\times101$, and $101\times101$ sizes, respectively. Matrix $\vb{V}$ consists of the singular vectors, and the entries of $\vb{\Sigma}^2$ are the corresponding coefficients of the projection of $\vb{F}$ on to the principal vectors in $\vb{V}$.

\begin{figure}[!h]
    \includegraphics{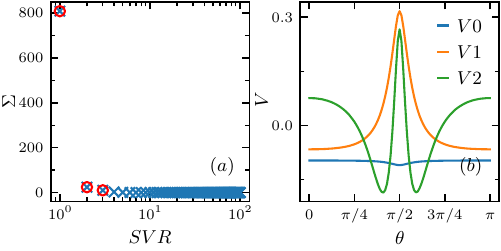}
    \caption{Singular value decomposition (SVD) of the angular scattering correlation data set. (a) Singular value $\Sigma$ versus singular value rank (SVR), top 3 ranked values are highlighted with red circle. (b) Singular vectors corresponding to the first three singular values.}
    \label{fig:svd}
\end{figure}

Fig.~\ref{fig:svd}(a) shows the singular value versus it's rank, where the rapid decay of the singular value indicates the significance of the projection on to higher rank singular vectors quickly becomes negligible. Therefore, decomposition of the $g(\theta)$ into the top three singular vectors will provide a good approximation of the whole $g(\theta)$. Figure~\ref{fig:svd}(b) shows the singular vectors $(V1,V2,V3)$ corresponding to the first three singular values.

\begin{figure}[!h]
    \centering
    \includegraphics{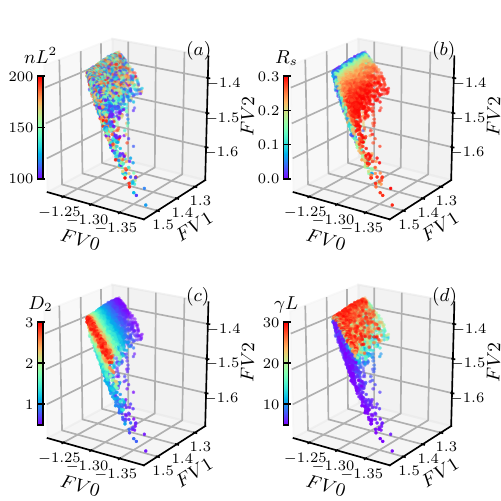}
    \caption{Distribution of the system parameters in the singular value space. (a) Number density $nL^2$. (b) Variation of particle size $R_s$. (c) Non-affine transformation $D_2$. (d) Shear rate $\gamma L$}
    \label{fig:SVD_projection}
\end{figure}

Projecting each $g(\theta)$ in the data set $\vb{F}$ on to the three singular vectors $(V1,V2,V3)$ yields three corresponding projection values $(FV0, FV1, FV2)$, which can be considered as a dimension reduction of the original $g(\theta)$. This converts each $g(\theta)$, as well as the system parameter $\vb{Y}=\left\{(nL^2,R_s,D_2,\gamma L)\right\}$ associated with it, in the data set $\vb{F}$ to a point in a three dimensional space spanned by the three projection values $(FV0, FV1, FV2)$. Fig.~\ref{fig:SVD_projection} shows the distribution of $(nL^2,R_s,D_2, \gamma L)$ in such a space. The distribution informs us the feasibility for mapping the features of $g(\theta)$ back to these system parameters. From the color distribution, we notice the values of $(R_s, D_2, \gamma L)$ are well spread out in the $(FV0, FV1, FV2)$ space, indicating a smooth and continuous mapping from the projection values of $g(\theta)$ to these corresponding system parameters. On the contrary, the distribution of number density $nL^2$ in the $(FV0, FV1, FV2)$ is rather random, implying it is not suitable for the inverted mapping. The inability to extract the number density information from our scattering correlation function is not surprising, as we are working in the limit of dilute dispersions where the number density of particles doesn't play a role in the microscopic structure or the rearrangement.

\subsection{Inference of the system parameters}

For the ML inversion of system parameters $(R_s,D_2,\gamma L)$ from angular scattering correlation function $g(\theta)$, we split the data set $\vb{F}=\left\{g(\theta)\right\}$ into two groups, a training set $\vb{F}_{train}=\left\{g(\theta)\right\}_{train}$ consists of $70\%$ of $\vb{F}$, and a test set $\vb{F}_{test}=\left\{g(\theta)\right\}_{test}$ consisting the rest of $30\%$. We use the training set to optimize the GPR, especially the hyperparameters $(l,\sigma)$ for each system parameter individually by maximize the log marginal likelihood using gradient descent\cite{williams2006gaussian}. We then use the trained GPR to predict the system parameters of the test set, and compare the GPR predicted system parameters with the ones actually used for the MC simulations.

\begin{figure}[!h]
    \centering
    \includegraphics{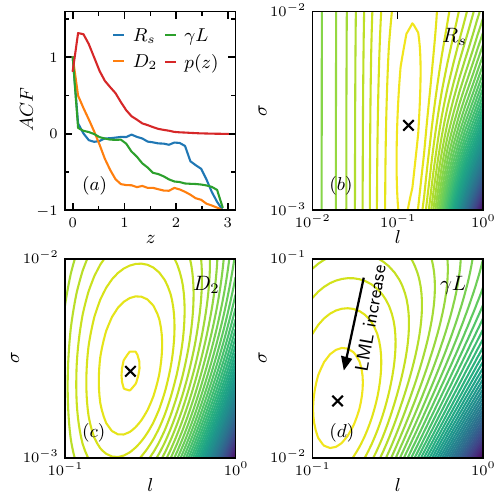}
    \caption{Determining the hyperparameters $l$ and $\sigma$ for system parameters. (a) Pair distance distribution function $p(z)$ of the data set $\vb{F}$ and autocorrelation function (ACF) for three system parameters. Log marginal likelihood (LML) of hyperparameters for system parameters: (b) $R_s$, (c) non-affine rearrangement $D_2$, and (d) affine shear strain $\gamma L$.}
    \label{fig:LML}
\end{figure}

Figure~\ref{fig:LML} shows the determination of log marginal likelihood contour in the $(l,\sigma)$ space for each system parameters $(R_s,D_2,\gamma L)$. To gauge the proper range of the $l$ for each system parameters, as shown in Figure~\ref{fig:LML}(a), we firstly analyze the PDDF of the $g(\theta)\in \vb{F}$ and then investigate the ACF for $(R_s, D_2, \gamma L)$, which gives us a rough range in which we search for the optimized $l$. Consequently, the log marginal likelihood contours are shown in Figure~\ref{fig:LML}(b)-(d), and the values of optimized $(l,\sigma)$ are shown in Table~\ref{tab: hyperparameter}.

\begin{table}[!hbt]
    \centering
    \begin{tabular}{|*{3}{p{0.25\linewidth}|}}
        \hline
                 & $l$ & $\sigma$ \\ \hline
        $R_s$ &  \num{1.338e-01} & \num{2.673e-03} \\ \hline  
        $D_2$ &  \num{2.419e-01} & \num{2.717e-03}  \\ \hline      
        $\gamma L$ & \num{1.405e-01} & \num{1.927e-02} \\ \hline 
    \end{tabular}
    \caption{Optimized hyperparameters for each feature, obtained from maximum log marginal likelihood.}
    \label{tab: hyperparameter}
\end{table}

\begin{figure*}[tbh]
    \centering
    \includegraphics{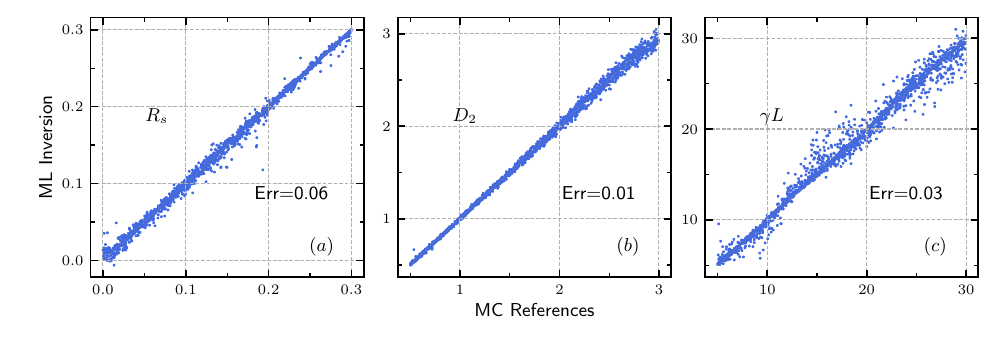}
    \caption{Comparison between the system parameters extracted from the angular scattering correlation function $g(\theta)$ using the GPR and their corresponding MC reference used for generating scattering data. Averaged relative error $Err$ is indicated in each plots. (a) Variation of particles size $R_s$. (b) Non-affine rearrangement $D_2$. (c) Affine shear strain $\gamma L$.}
    \label{fig:GPR}
\end{figure*}

We apply the trained GPR with the optimized hyperparameters on the test set $F_{test}$ to infer ML inverted system parameters $(R_s, D_2, \gamma L)$, and compare the inferred results with MC references. Fig.~\ref{fig:GPR} shows the comparison of system parameters $(R_s, D_2, \gamma L)$, where almost all of the data points lay around the diagonal line indicating a good estimation. For each system parameter $\mu$, the relative error between MC reference $\mu_{MC}$ and ML inversion $\mu_{ML}$ is estimated by $Err = \left<|\mu_{MC}-\mu_{ML}|/max(\mu_{MC},\mu_{ML}) \right>$, where $\left<\dots\right>$ here is average over all data points.  The relative error is labeled for each system parameter in each panel of Fig.~\ref{fig:GPR} and shows a very high precision: 1\% for $D_2$, and 3\% for $\gamma L$, and 6\% for $R_s$. The precised quantification of system parameters demonstrates the power of our ML approach for analyzing and extracting microscopic rearrangement from coherent scattering data.

\section{Summary}
\label{sec:summary}
In this work, we present a machine learning-informed analysis framework that successfully recovers the polydispersity and microscopic rearrangements, including both affine simple shear and non-affine transformation, with high precision from the correlation function $g(\vb{q})$ of coherent scattering intensity of the dilute dispersions of spherical particles. 

Our simulated colloidal systems and scattering intensity aim to mirror real synchrotron scattering setups, including the beam size, particle size, and the range of the detectable wave vector, therefore, our SVD features and GPR models can be easily compared and adopted to analyze real experimental data. The direction of affine shear is not necessarily always in $x$-direction for experimental data, however, the direction is easy to identify by the high correlation strip in the $g(\vb{q})$ pattern, one can rotate the scattering data before applying our model. In case that the experimental scattering data is taken on a different grid of wave vector $\vb{q}$, interpolated data can be utilized to feed into our GPR model. Alternatively, we can also use generative model such that Kolmogorov-Arnold Networks\cite{liu2024kan,liu2024kan2} to obtain the $I(\vb{q})$ as a continues function of the system parameters and use it to directly fit the experimental data.

Further, through averaging $I(\vb{q})$ and $g(\vb{q})$ over $2\times 10^4$ samples of particle configurations for each transformation $\Gamma$, the features of angular correlation function and trained hyperparameters we obtained are independent of any specific realization of the particle positions. They can be applied to analyze the correlation function of two instances of scattering pattern without any time average, and this can greatly improve the time resolution for studies of the microscopic rearrangements involved in non-steady and transient dynamics. Moreover, similar method can be deployed to study the microscopic rearrangement of disordered colloidal systems, like glasses and gels, where machine learning-assisted quantification methods have a great potential to overcome the challenges imposed by the out-of-equilibrium nature~\cite{schoenholz2016structural, horwath2024ai}.

\begin{acknowledgments}
This research was performed at the Spallation Neutron Source, which is DOE Office of Science User Facilities operated by Oak Ridge National Laboratory. This research was sponsored by the Laboratory Directed Research and Development Program of Oak Ridge National Laboratory, managed by UT-Battelle, LLC, for the U. S. Department of Energy. The ML aspects were supported by by the U.S. Department of Energy Office of Science, Office of Basic Energy Sciences Data, Artificial Intelligence and Machine Learning at DOE Scientific User Facilities Program under Award Number 34532. Monte Carlo simulations and computations used resources of the Oak Ridge Leadership Computing Facility, which is supported by the DOE Office of Science under Contract DE-AC05-00OR22725 and resources of the National Energy Research Scientific Computing Center, which is supported by the Office of Science of the U.S. Department of Energy under Contract No. DE-AC02-05CH11231. Y. C. acknowledges the support by the NSF through Grant No. DMR-2003659 and by the NSF Penn Materials Research Science and Engineering Center through Grant No. DMR-2309043, and thanks R. L. Leheny and A. G. Yodh for helpful discussion.
\end{acknowledgments}


\bibliography{apssamp}

\end{document}